\documentclass{aastex63}

\usepackage{color}

\shorttitle{Energy Partition in Magnetic Reconnection}
\shortauthors{Hoshino}
\begin{document}
\title{Energy Partition of Thermal and Nonthermal Particles in Magnetic Reconnection}
\correspondingauthor{Masahiro Hoshino}
\email{hoshino@eps.s.u-tokyo.ac.jp}
\author[0000-0002-1818-9927]{Masahiro Hoshino}
\affil{Department of Earth and Planetary Science,The University of Tokyo, Tokyo 113-0033, Japan}
\begin{abstract}\
Magnetic reconnection has long been known to be the most important mechanism as quick conversion of magnetic field energy into plasma kinetic energy. In addition, energy dissipation by reconnection has gained attention not only as a plasma heating mechanism, but also as a plasma mechanism for accelerating nonthermal particles. However, the energy partitioning of thermal and nonthermal plasmas during magnetic reconnection is not understood. Here, we studied energy partition as a function of plasma sheet temperature and guide magnetic field. 
In relativistic reconnection with anti-parallel magnetic field or weak guide magnetic field, it was found that the nonthermal energy density can occupy more than $90 \%$ of the total kinetic plasma energy density, but strengthening the guide magnetic field suppresses the efficiency of the nonthermal particle acceleration. 
In nonrelativistic reconnection for anti-parallel magnetic field, most dissipated magnetic field energy is converted into thermal plasma heating.  For a weak guide magnetic field with a moderate value, however, the nonthermal particle acceleration efficiency was enhanced, but strengthening the guide-field beyond the moderate value suppresses the efficiency. 
\end{abstract}


\section{Introduction}
Magnetic reconnection is known to be the most important mechanism not only for plasma thermalization up to the equivalent temperature of the Alfv\'{e}n velocity, but also for accelerating nonthermal particles whose energies exceed their thermal energies \citep[for example,][]{Birn07,Zweibel09,Hoshino12,Uzdensky16,Blandford17}. Recently, nonthermal particle acceleration during reconnection has been investigated through satellite observations of the Earth's magnetosphere and the solar atmosphere \citep[for example,][]{Oieroset2002,Lin2003}. It has been revealed that reconnection in the heliosphere with nonrelativistic plasmas can generate nonthermal high-energy particles with a non-negligible fraction compared to thermal plasma heating. It is also found that nonthermal particles can be characterized by a power-law spectrum, whose spectral indices are not necessarily hard compared with the cosmic ray observations, which are believed to be generated by the standard diffusive shock acceleration \citep[for example,][]{Bell78,Blandford78}. As magnetic reconnection is believed to be an ubiquitous phenomenon in the plasma universe, many high-energy astrophysical phenomena can occur, such as those in pulsar magnetosphere, accretion disks, and magnetar \citep[for example,][]{Remillard06,Done07,Madejski16,Kirk04,Lyutikov03}. The nonthermal energy spectra observed from high-energy astroplasma objects often indicate a hard power-law spectrum, which cannot be explained by standard shock diffusive acceleration.
Magnetic reconnection in relativistic plasmas has also been investigated as a possible alternative acceleration process. It has been found that relativistic reconnection whose Alfv\'{e}n velocity is close to the speed of light can effectively generate a nonthermal energy spectrum with a harder power-law spectrum than that expected in the standard diffusive shock acceleration \citep[for example,][]{Zenitani01,Jaroschek04}. Since then, regardless of whether the plasma is nonrelativistic or relativistic, magnetic reconnection has gained attention as a mechanism of nonthermal particle acceleration in various astrophysical sources.

To understand comprehensively the nonthermal particle acceleration in magnetic reconnection, a number of particle-in-cell (PIC) simulation studies have been extensively conducted \citep[for example,][]{Zenitani05a,Zenitani05b,Zenitani07,Zenitani08,Jaroschek09,Liu11,Sironi11,Sironi14,Hoshino01,Hoshino15,Hoshino18,Bessho12,Cerutti12a,Cerutti12b,Cerutti13,Cerutti14,Guo14}. Although previous studies have identified the formation of a harder power-law spectrum for a stronger magnetic field with a faster Alfv\'{e}n velocity, an important question of the energy partition between thermal and nonthermal plasmas has not been addressed.
Recently, we discussed energy partitioning during magnetic reconnection for an anti-parallel magnetic field topology by \cite{Hoshino22} (Paper I). The production of nonthermal particles increases with increasing Alfv\'{e}n speed for magnetic reconnection. In the relativistic reconnection, where the Alfv\'{e}n speed is close to the speed of light, it was found that the nonthermal energy density can occupy more than $90 \%$ of the total kinetic plasma energy density. However, most dissipated magnetic field energy can be converted into thermal plasma heating in nonrelativistic reconnection.

The effect of guide field is believed to be another important factor for controlling particle acceleration. Here, we study the energy partition between thermal and non-thermal populations by considering both the plasma sheet temperature and the magnitude of guide magnetic field.
Thus, we focus on the time and spatial evolution of the energy spectrum during reconnection. The energy spectrum obtained through reconnection simulation, in general, comprises thermal and nonthermal populations. A model fitting of the energy spectrum through combination of Maxwellian and kappa distributions \citep[for example, ][]{Vasyliunas68} is performed in the same manner as in Paper I. We quantify the thermal and nonthermal populations as functions of the guide magnetic field and initial plasma temperature, both in nonrelativistic and relativistic regimes.
We prove that thermal plasma heating dominates over nonthermal particle acceleration for nonrelativistic reconnection, and that the efficiency of nonthermal particle acceleration is enhanced for relativistic reconnection. For nonrelativistic plasma, a weak magnetic field slightly enhances the efficiency of nonthermal production. However, for both nonrelativistic and relativistic reconnections, the efficiency for stronger guide-field reconnection is suppressed.

\section{Overview of Simulation Study}
In our simulation study, we use the same numerical setting as in our previous study (Paper I), except for the guide magnetic field, and study the time evolution of reconnection using a two-dimensional PIC simulation. We assume a periodic boundary in the $x$-direction for $x=-215 \lambda$ and $x=215 \lambda$, and conducting walls for the upper and lower boundaries at $y= \pm 215 \lambda$, where $\lambda$ is the thickness of the initial plasma sheet. The total system size is $L_x \times L_y = 430 \lambda \times 430 \lambda$ and the computational grid size is $5376 \times 5376$. The total number of particles was $8.4 \times 10^{10}$, and the average number of particles per grid size is $2.9 \times 10^3$.  These values are used to accurately calculate the magnetic energy dissipation process in the high-$\beta$ plasma sheet.

\begin{figure}
\begin{center}
\includegraphics[width=9cm]{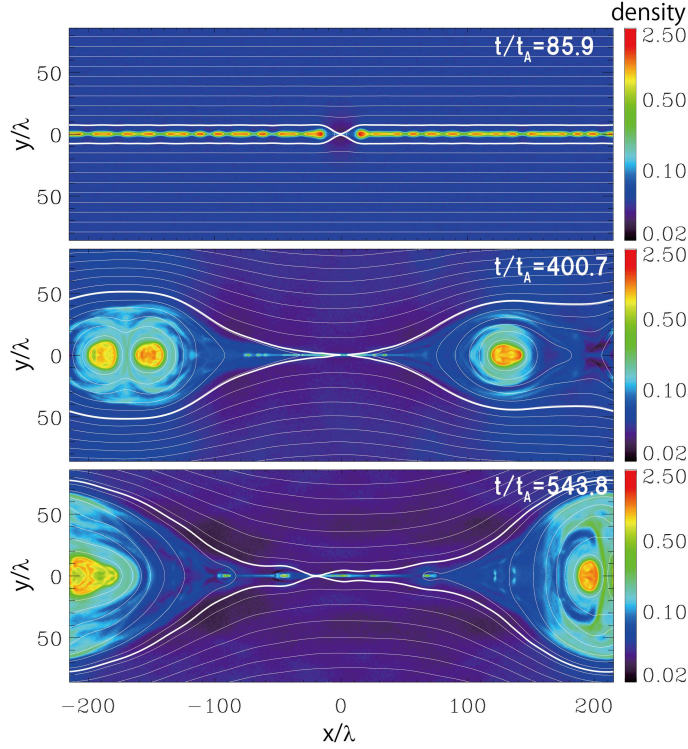}
\caption{Time evolution of a pair plasma density during collisionless magnetic reconnection obtained through PIC simulation
for a plasma sheet temperature $T_0/mc^2=1$ and a guide magnetic field $B_G/B_0=1/2$. 
The top panel ($t/\tau_A=85.9$) shows the early stage of reconnection, and the main X-type neutral point at the center of the simulation box and other smaller tearing islands were formed. The middle panel ($t/\tau_A=400.7$) was a nonlinear stage formed after many small-scale plasmoids coalesced, forming larger plasmoids. The bottom panel ($t/\tau_A=543.8$) was nearly the final stage when two large plasmoids coalesced into one large plasmoid in the system.
The thick white lines indicate the magnetic field lines that passed through the X-type neutral line, that is, the separatrix between the downstream region and the upstream plasma sheet. The color bars in the right-hand side indicate the plasma density normalized by the initial plasma sheet density in the logarithmic scale.}
\label{fig:time_evolution}
\end{center}
\end{figure}

For simplicity, we adopt the Harris solution \citep{Harris62} for the pair plasma with the same mass $m$ and uniform temperature $T_0$ in space and focused on the energy partition of an idealized magnetic reconnection in a collisionless plasma system. 
The magnetic field ${\bf B}(y)$ and plasma density $N(y)$ are expressed as
\begin{equation}
  {\bf B}(y) = B_0 \tanh( y/\lambda) {\bf e_x} + B_G {\bf e_z},
\end{equation}
and
\begin{equation}
N(y) = N_0 \cosh^{-2}(y/\lambda) + N_{\rm b},
\label{eq:Harris_den}
\end{equation}
respectively, where $B_G$ is the uniform guide magnetic field perpendicular to the reconnection plane, and $N_{\rm b}$ is the background uniform plasma density adopted to demonstrate continuous plasma injection from the outside plasma sheet to the reconnection downstream (exhaust). The background plasma density $N_{\rm b}$ is set as $5 \%$ of the maximum plasma density $N(0)$ at $y=0$. The relationship between the pressure balance $B_0^2/ 8 \pi = 2 N_0 T_0$ and force balance $2 T_0/\lambda = |e| u_d B_0/c$ is satisfied, where $u_d$ is the drift velocity, and the initial electric field ${\bf E}$ is zero.

We investigate the energy partition between the thermal and nonthermal populations by changing the magnitude of the guide magnetic field $B_G$ and the plasma temperature $T_0$. We investigate 
\begin{equation}
  B_G/B_0=0, ~1/8, ~1/4, ~1/2  {\rm~and~} 1,
\end{equation}
and
\begin{equation}
  T_0/mc^2 = 10^{n-2} \quad {\rm with} \quad n=0,~1,~2 {\rm~and~} 3.
\end{equation}
These temperatures correspond to the magnetization parameter $\sigma = B_0^2/(8 \pi N_0 mc^2)= 2 \times 10^{n-2}$ if we use the number density for the value of the Harris plasma sheet $N_0$, owing to the pressure balance between the gas pressure inside the plasma sheet and the magnetic pressure outside the plasma sheet.
The Alfv\'{e}n speed $v_A$ is expressed as $v_A =c\sqrt{\sigma/(1+\sigma)}$
in the cold plasma limit, where the density and magnetic field are used for the values of the central Harris plasma sheet and outside plasma sheets, respectively.
The inertia length $c/\omega_p$ in the plasma sheet is about $3.3 \sim 7.9$ grid cell for $T_0/mc^2=10 \sim 10^{-2}$. 
We maintain the ratio of the gyro-radius $r_g$ for the Harris magnetic field $B_0$ and the thickness of the plasma sheet $\lambda$ for all simulation runs and set $r_g/\lambda=0.45$, where $r_g = mc^2 \sqrt{\gamma_{th}^2-1}/(e B_0)$. The gyro-radius $r_g$ decreases with an increase in the guide magnetic field.
For simplicity, we assume $(\gamma_{th}-1)mc^2/T_0=1$; then, the drift speed $u_d$ is calculated using $u_d/c=2 (r_g/\lambda)/\sqrt{1+2 mc^2/T_0}$. 
We add a small initial perturbation for the vector potential $\delta A_z$ to the Harris equilibrium to initiate the formation of an X-type reconnection at the center of the simulation box. The initial amplitude of the reconnected magnetic field is set to $\max(B_y/B_0)=10^{-2}$ in the neutral sheet.

\begin{figure*}
\includegraphics[width=18cm]{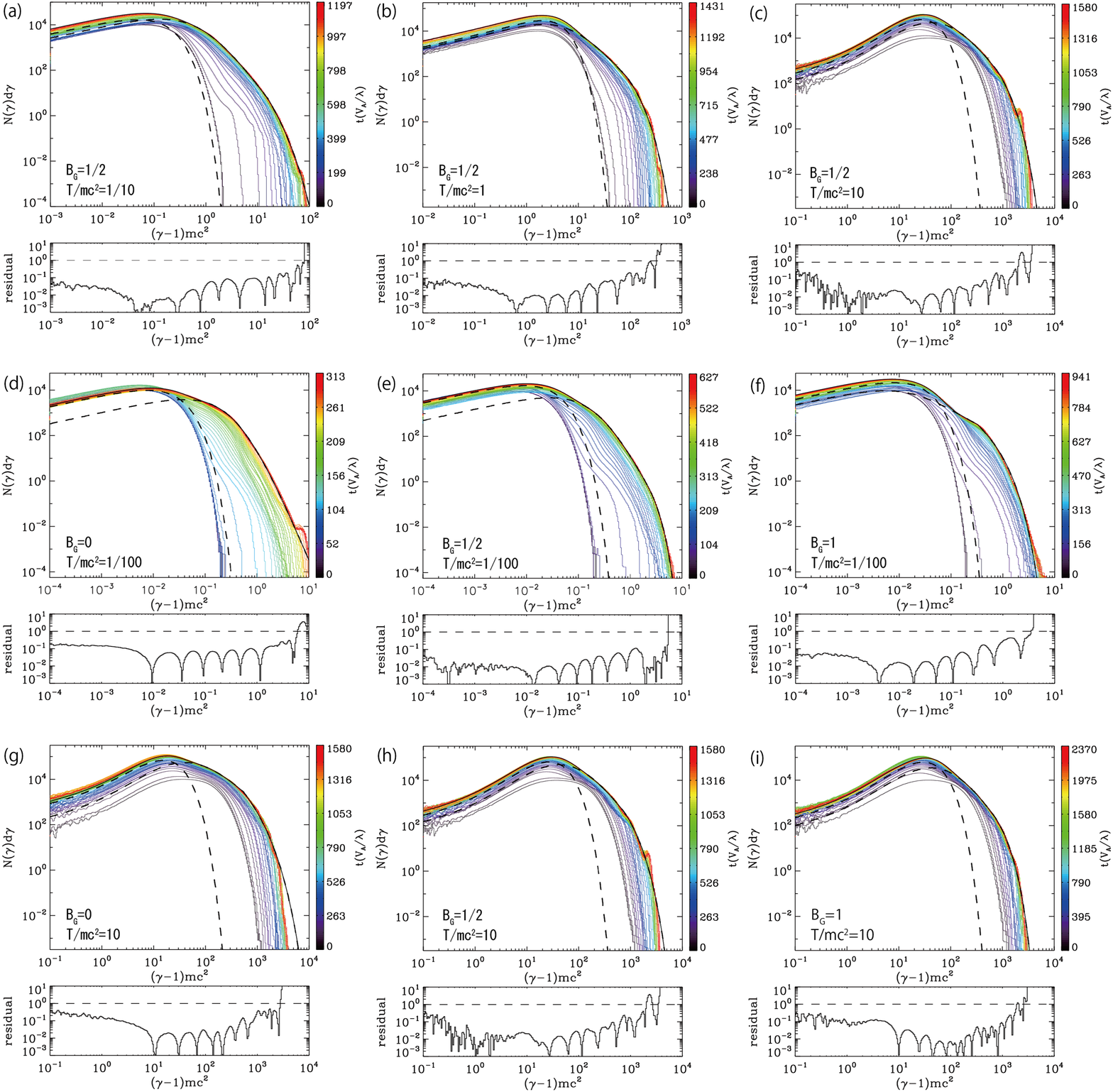}
\caption{Energy spectra for the downstream region for the same guide magnetic field (a) $(B_G/B_0,T_0/mc^2)=(0.5,0.1)$, (b) $(0.5,1)$, (c) $(0.5,10)$, for the same cold plasma temperature (d) $(0,0.01)$, (e) $(0.5,0.01)$, (f) $(1,0.01)$, and for the same hot plasma temperature (g) $(0,10)$, (h) $(0.5,10)$, (i) $(1,10)$.  The color lines indicate the time evolution of the spectra, whose time stages are indicated in the right-hand side bar. The dark blue line indicates the initial state with the Maxwell distribution, and the red line indicates the final stage when a large plasmoid was formed in the simulation box. The thick solid lines represent the model fitting for the composed function of the Maxwellian and kappa distributions $N_{{\rm M}+\kappa}(\gamma)$. The two black dashed lines represent the Maxwellian part of the fitting result $N_{\rm M}(\gamma)$ in the lower energy regime and the kappa distribution parts $N_{\kappa}(\gamma)$ in the higher energy regime. The bottom panels indicate the residual of the model fitting, which show the difference between the model fitting and simulation result. The dashed lines in the bottom plot may represent an acceptable residual as reference.}
\label{fig:spectra}
\end{figure*}

Figure \ref{fig:time_evolution} shows three time stages of the reconnection structure obtained from the particle-in-cell simulation for a temperature of $T_0/mc^2=1$ and guide magnetic field of $B_G/B_0=1/2$. The pair plasma density normalized by the initial central plasma sheet density and the magnetic field lines are indicated by the colored contour and white lines, respectively. The thick white lines indicate the separatrix of the magnetic reconnection for the X-type neutral point formed around the center $x/\lambda \sim 0$, that is, the most recently reconnected magnetic-field lines.
These lines indicating the separatrix are obtained from the contour lines of the vector potential $A_z$, which has the minimum value in the neutral line $y=0$. Only part of the simulation system is illustrated.

The top panel shows the early time stage at $t/\tau_A=85.9$ when the reconnection started at the center, where $\tau_A=\lambda/v_A$ is the Alfv\'{e}n transit time, and many small-scale tearing islands formed along $y=0$. The middle panel at $t/\tau_A=400.7$ indicates the time stage after the small tearing islands coalesced, and the two medium-sized plasmoids merged, forming a larger plasmoid at $x/\lambda \sim -170$. Because we imposed an initial small perturbation $\delta A_z$ at $x/\lambda=0$, the most prominent reconnection occurred at the center. The bottom panel at $t/\tau_A=543.8$ is the approximate final stage of the active reconnection in our periodic system after the two large plasmoids merged, forming one large magnetic island. In addition, several plasmoids were generated in the elongated magnetic diffusion region at the center. A multiscale structure consisted of small-scale plasmoids and large-scale reconnection island are observed. The time evolution shows the general behavior of magnetic reconnection, regardless of the guide magnetic field.

\section{Model Fitting of Energy Spectrum}
Because our objective is to understand the energy partition between thermal and nonthermal populations during reconnection, we investigate the energy spectrum downstream of the separatrix lines shown in Figure \ref{fig:time_evolution}. Figure \ref{fig:spectra} illustrates the time evolution of nine energy spectra: The top three panels are the comparison of the same guide magnetic field $B_G/B_0=1/2$ for (a) $T_0/mc^2=10^{-1}$, (b) $T_0/mc^2=1$, and (c) $T_0/mc^2=10$. The middle three panels are the comparison of the same cold plasma temperature $T/mc^2=1/00$ for (d) $B_G/B_0=0$, (e) $B_G/B_0=1/2$, and (f) $B_G/B_0=1$.  The bottom three panels are that of the hot plasma temperature $T_0/mc^2=10$ for (g) $B_G/B_0=0$ (h), $B_G/B_0=1/2$, and (i) $B_G/B_0=1$. 
These spectra were obtained by integrating particle energy over the reconnection region sandwiched by the separatrix, that is, the thick white lines shown in Figure \ref{fig:time_evolution}. Note that when the width of the separatrix is smaller than the thickness of the plasma sheet $\lambda$, integration is conducted for the plasma sheet with size $\lambda$.
The horizontal axis represents the particle kinetic energy $(\gamma -1)mc^2$ and the vertical axis represents the number density $N(\gamma) d\gamma$, where $\gamma=1/\sqrt{1-(v/c)^2}$.
The vertical scales of $N(\gamma)d\gamma$ are shown in an arbitrary unit, but the ratio of the top to the bottom is adjusted to be a nine order-of-magnitude.  In the horizontal axis, the number of particle is counted by logarithmic binning, so the one-count level is below the spectra shown here. 
The colored lines indicate the time evolution of the energy spectra, whose time stages are indicated by the colored bars on the right-hand side. The bluish and reddish colors correspond to the earlier and later stages, respectively. Because we assumed a uniform plasma temperature in space for both the $\cosh$-type Harris density and background populations in Equation (\ref{eq:Harris_den}), the dark blue spectra in Figure \ref{fig:spectra} indicate the initial plasma state with the initial Maxwellian distribution functions. However, for the relativistic case shown in Figures 2c, 2g, 2h, and 2i, the one-dimensional energy spectra were modified by the effect of a high-speed drift velocity of $u_d/c \sim 0.82$.  

Over time, we clearly observe that the Maxwellian plasmas are gradually heated owing to the reconnection heating process, and the high-energy components were further accelerated to form a nonthermal population.
The behavior of these plasma heating and particle acceleration is basically same as many previous simulation studies \citep[for example,][]{Zenitani01,Hoshino01,Jaroschek04,Drake06,Liu11,Sironi11,Bessho12,Cerutti12a,Guo14,Haggerty15,French22}. 
Through examining of the energy interval of the time evolution of the energy spectra, it was found that the energy intervals were wider in the early acceleration stage compared with those in the later stage, suggesting that rapid energy gain occurs in the early stage.

Because the spectra in the final stage, indicated by the red curve, are approximately stable, we investigate the spectral behavior of the red curve by a chi-square fitting for a model function. As discussed in Paper I, the final-stage spectra can be well fitted by a composed model spectrum $N_{M+\kappa}(\gamma)$ of a Maxwell distribution function $N_M(\gamma)$ and a kappa distribution function $N_{\kappa}(\gamma)$, as described below.
\begin{equation}
N_{M+\kappa}(\gamma) = N_M(\gamma) +N_{\kappa}(\gamma),
\label{eq:modelfunction}
\end{equation}
where
\begin{equation}
N_M(\gamma) =n_M \gamma \sqrt{\gamma^2-1} \exp(-\frac{\gamma-1}{T_M/mc^2}),
\end{equation}
and
\begin{equation}
N_{\kappa}(\gamma) =n_{\kappa} \gamma \sqrt{\gamma^2-1}
\left(1 + \frac{\gamma-1}{\kappa T_{\kappa}/mc^2} \right)^{-(1+\kappa)} f_{cut}(\gamma).
\label{eq:kappa_distribution}
\end{equation}
For simplicity, a three-dimensional distribution function was projected into a one-dimensional distribution in the simulation frame as a function of the particle energy $\gamma$.
The kappa distribution comprises a thermal Maxwellian distribution at low energies and a nonthermal population approximated by a power-law function at high energies \citep[for example, ][]{Vasyliunas68}.
Note that a kappa distribution function approaches a Maxwellian distribution as $\kappa \to \infty$, as described below.
\begin{eqnarray}
\lim_{\kappa \to \infty} \left(1 + \frac{\gamma-1}{\kappa T_{\kappa}/mc^2} \right)^{-(\kappa+1)} \simeq
    \exp \left( - \frac{\gamma -1}{T_{\kappa}/mc^2} \right).
\end{eqnarray}
When we analyze the model fitting, we add a high-energy cutoff function $f_{cut}$ to represent a possible high-energy cutoff in the simulation data \citep{Oka15,Werner16,Petropoulou18}, which is described as

\begin{equation}
f_{cut}(\gamma) = \left\{
\begin{array}{@{\,}ll}
1 & \mbox{for $\gamma \le \gamma_{cut}$ }, \\
\exp\left(-(\gamma -\gamma_{cut})/\gamma_{cut} \right) & \mbox{for $\gamma > \gamma_{cut}$ }.
\end{array}
\right.
\end{equation}
A high-energy cutoff may originate from the finite time evolution of reconnection under a limited system size \citep{Petropoulou18}.
Although high-energy cutoff is important in understanding the total energy budget of nonthermal population, the dominant contribution of the nonthermal energy density comes from the low-energy population as long as $\kappa > 3$.
The high energies of $N_{\kappa}$ was approximated using a power-law function $N_{\kappa} \propto \gamma^{-\kappa+1}$ with a power-law index $s=\kappa-1$ for $\gamma\gg T_{\kappa}/mc^2$.

In the nine panels of $N(\gamma) d\gamma$ shown in Figure \ref{fig:spectra}, the thick solid lines are the model fitting curves for $N_{M+\kappa}$ in the final stage, whereas the two dashed lines are the fitting curves of $N_M$ at lower energies and $N_{\kappa}$ at higher energies. It was found that these nine simulation spectra in the downstream region were approximated by the model function consisting of the Maxwellian and kappa distributions $N_{{\rm M}+\kappa}(\gamma)$.  It was also confirmed that all simulation cases of $T_0/mc^2 = 10^{n-2}$ with $n=0,1,2$ and $3$, and $B_G/B_0=0,1/8,1/4,1/2$ and $1$ exhibited a similar good model fitting.

The bottom three panels indicate the error of the model fitting, described as
\begin{eqnarray}
{\rm residual}(\gamma) = |N_{data}(\gamma) - N_{M+\kappa}(\gamma) |/
{\rm min}(N_{data}(\gamma),N_{M+\kappa}(\gamma)),
\end{eqnarray}
where ${\rm min()}$ denotes a function that returns the minimum element from $N_{data}$ and $N_{M+\kappa}$. If $\frac{1}{2}N_{data} < N_{M+\kappa} < 2 N_{data}$, the error is less than $1$, as depicted by the dashed line in the bottom panel of Figure \ref{fig:spectra}.
We observed that the residuals in wide energy ranges were less than $10^{-1}$, and the model fitting of $N_{M+\kappa}$ described the energy spectra obtained in the simulation runs well.

\section{Energy Partition based on Model Fitting}
\begin{figure*}
\includegraphics[width=18cm]{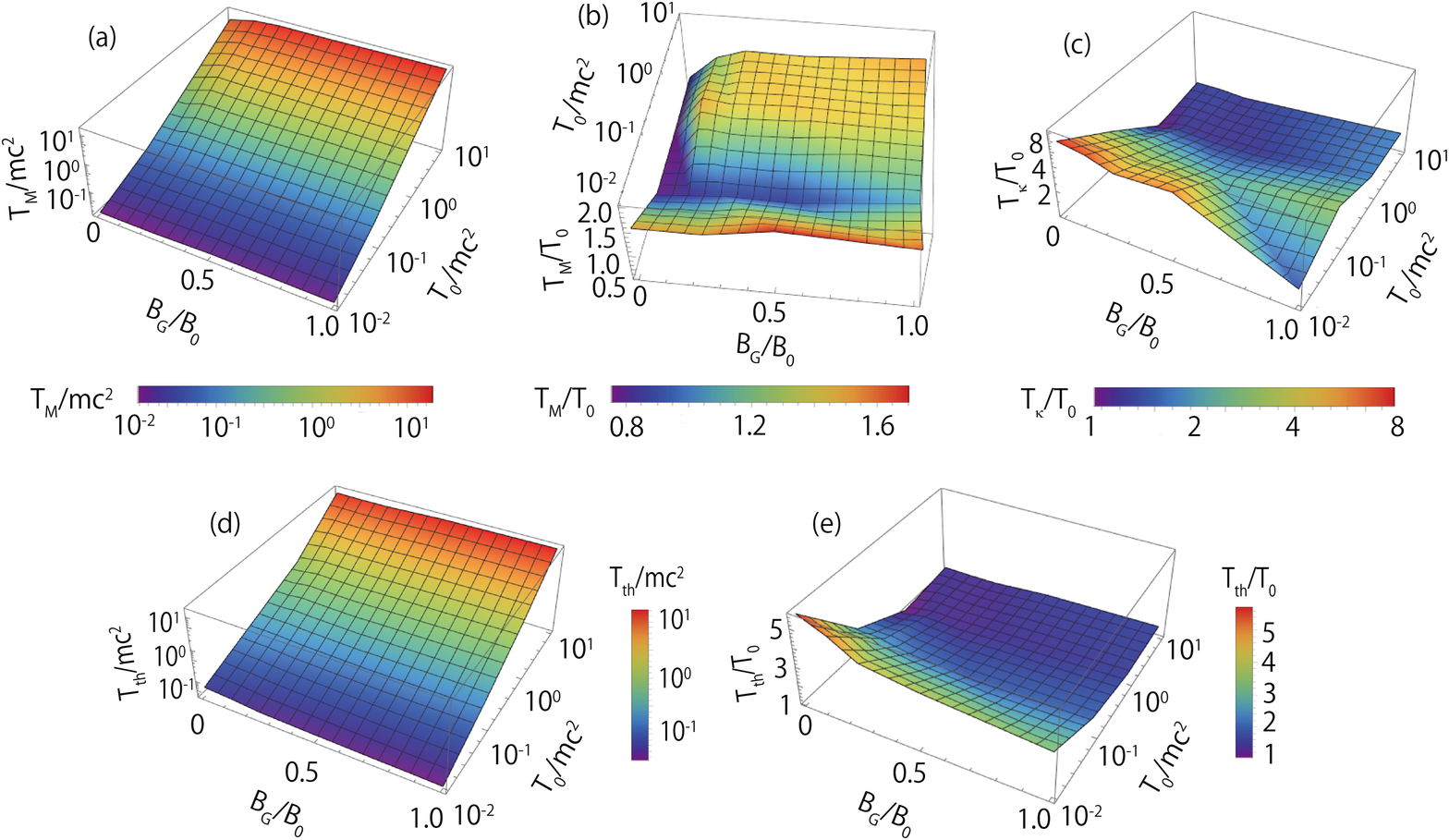}
\caption{Model fitting results as functions of the initial plasma temperatures $T_0/mc^2=10^{-2} \sim 10^1$ and the guide magnetic fields of $B_G/B_0=0 \sim 1$. The top three panels show (a) the Maxwellian temperatures $T_M/mc^2$ normalized by the rest-mass energy $mc^2$, (b) the Maxwellian temperature $T_M/T_0$ normalized by the initial plasma temperature $T_0$, and (c) the temperature of the kappa distribution function $T_{\kappa}/T_0$ normalized by the initial plasma temperature $T_0$. 
The bottom two panels are (d) the average thermal temperatures $T_{th}/mc^2$ of the Maxwellian function $T_M$ and the kappa function $T_{\kappa}$ normalized by the rest-mass energy $mc^2$ and (e) the average temperatures $T_{th}/T_0$ normalized by the initial background temperature $T_0$.}
\label{fig:thermal}
\end{figure*}

Figure \ref{fig:thermal} shows the model fitting results for (a) the Maxwellian temperature ($T_M/mc^2$) normalized by the rest-mass energy $mc^2$, (b) the same Maxwellian temperature ($T_M/T_0$) normalized by the initial plasma temperature $T_0$, and (c) the normalized temperature of $\kappa$ distribution function ($T_{\kappa}/T_0$). We found that the Maxwellian temperature $T_M/mc^2$ in Panel (a) was almost the same as the initial plasma temperature ($T_0/mc^2$), and that temperature $T_M/T_0$ normalized by the initial plasma temperature $T_0$ in Panel (b) ranged from 0.8 to 1.6.
Note that $T_M < T_0$ may be caused by the slow mode expansion waves propagating outward form the magnetic diffusion region \citep{Hoshino18}.  
The normalized thermal temperature of $\kappa$ distribution function $T_{\kappa}/T_0$ is shown in Panel (c).
In general, the normalized temperature $T_{\kappa}/T_0$ also exhibited the order of the initial temperature; however, we observed that $T_{\kappa}$ was several times higher than $T_M$ for the regime around the cold background temperature $T_0/mc^2 < 10^{-1}$ and weak guide field $B_G/B_0 < 0.5$. At $T_0/mc^2=10^{-2}$ and $B_G=0$, the normalized temperature $T_{\kappa}/T_0$ was 7.6, and the thermal plasmas for the $\kappa$ distribution were significantly heated.

Although we show the individual thermal temperatures of the Maxwellian and $\kappa$ functions in Panels (b) and (c), it is important to know the average thermal temperature of the Maxwellian and $\kappa$ functions, which is described as
\begin{equation}
  T_{th}=(n_M T_M + n_{\kappa} T_{\kappa})/(n_M + n_{\kappa}).
\end{equation}
The left-hand Panel (d) in Figure \ref{fig:thermal} shows the average thermal temperature normalized by the rest-mass energy $mc^2$, whereas the right-hand Panel (e) shows the temperature normalized by the initial background temperature $T_0$.
As expected, $T_{th}/mc^2$ in Panel (d) was almost the same as the initial plasma temperature $T_0/mc^2$ ; however, we observed some unique characteristics in the normalized temperature $T_{th}/T_0$ in Panel (e).  
Although there were hollow and bump profiles in Panels (b) and (c), these profiles were smeared for the average temperature $T_{th}/T_0$ in Panel (e). The average temperature was almost constant over a wide range of parameters, except for $T_0/mc^2 \sim 10^{-2}$. The most important result is the significant heating observed for the nonrelativistic reconnection, in which the heating slightly decreases with increasing guide magnetic field.  

\begin{figure*}
\includegraphics[width=18cm]{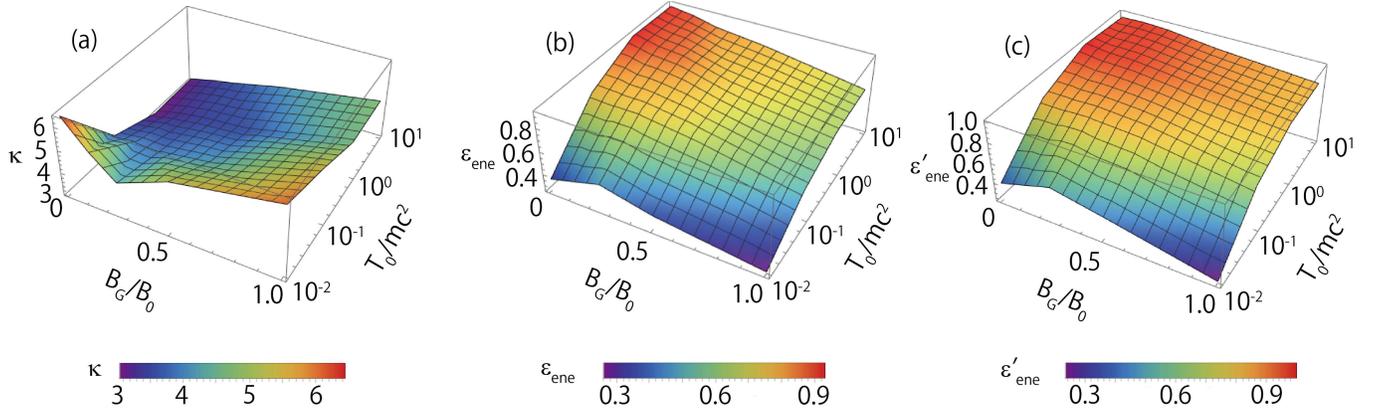}
\caption{Model fitting results of (a) $\kappa$ index, (b) efficiency of nonthermal particles against thermal plasma, and (c) efficiency of nonthermal particles against heated thermal plasma as a function of the initial plasma temperature $T_0/mc^2$ and guide magnetic field $B_G/B_0$.}
\label{fig:nonthermal}
\end{figure*}

For the nonthermal particle acceleration, Figure \ref{fig:nonthermal} shows (a) the dependence of $\kappa$ index (left) and (b) the fraction of the nonthermal energy density $\varepsilon_{\rm ene}$ (middle) as a function of the plasma temperature ($T_0/mc^2$) and guide field ($B_G/B_0$), where $\varepsilon_{\rm ene}$ is defined as
\begin{eqnarray}
  \varepsilon_{\rm ene}=
  \int_1^{\infty} (\gamma -1) (N_{\kappa}(\gamma)-N_{\kappa}^{\rm M}(\gamma)) d \gamma /
       \int_1^{\infty} (\gamma -1) N_{{\rm M}+\kappa}(\gamma) d \gamma,
\label{eq:efficiency}
\end{eqnarray}
where $N_{\kappa}^{\rm M}(\gamma)$ represents the portion of the Maxwellian distribution function in the $\kappa$ distribution function, that is, the $\kappa$ value was replaced by $\kappa=\infty$, while the other fitted parameters remained the same. Therefore, the numerator of $N_{\kappa}(\gamma)-N_{\kappa}^{\rm M}(\gamma)$ corresponds only to the nonthermal power-law component.

When $B_G=0$, the result is the same as that discussed in Paper I. The $\kappa$ index decreases with an increase in the plasma temperature from nonrelativistic cold plasma to relativistic hot plasma. For a finite guide magnetic field of $B_G \ne 0$, we observe a similar dependence, such that the $\kappa$ index decreases with increasing plasma temperature, except in the nonrelativistic and weak guide field regime.  Based on the change in the profile along the guide magnetic field, the $\kappa$ index increases with the increasing magnitude of the guide magnetic field $B_G$.
Except for this anomaly around the weak guide magnetic field of $B_G/B_0=1/4$ and cold nonrelativistic plasma of $T_0/mc^2=10^{-2}$, the nonthermal spectrum becomes softer as the guide field became stronger.

Panel (b) shows the dependence of the nonthermal energy density $\varepsilon_{\rm ene}$ on the plasma temperature ($T_0/mc^2$ ) and guide field ($B_G/B_0$). As expected, there is a correlation between the hardness of nonthermal spectrum and the nonthermal energy density. When $B_G=0$, the result is the same as that discussed in Paper I. We found that $\varepsilon_{\rm ene}$ was more than 90\% for relativistic hot plasma, and decreased with decreasing plasma temperature; $\varepsilon_{\rm ene}$ was approximately 30\% for $T_0/mc^2=10^{-2}$. The nonthermal fraction is small for the nonrelativistic reconnection; however, $\varepsilon_{\rm ene} \sim 30 \%$ is not negligible for the thermal energy density; the nonthermal plasma plays an important role in the dynamic evolution of reconnection.

When we examine the variation of the efficiency $\varepsilon_{\rm ene}$ along the plasma temperature $T_0$ for a finite guide magnetic field $B_G/B_0 \ne 0$, the efficiency $\varepsilon_{\rm ene}$ increases with increasing plasma temperature. When we make the plasma temperature $T_0$ constant and examine the variation of the efficiency along the guide magnetic field $B_G$, the efficiency decreases with increasing guide magnetic field $B_G$, except for the cold plasma $T_0/mc^2 \sim 10^{-2}$ and $B_G/B_0 \sim 0.25$, where $\kappa$ index exhibits a relatively small value in Panel (a). Nonthermal particles with a hard energy spectrum have high efficiency for relativistic hot plasma with a weak guide field.

Because the thermal energy in the analysis of the middle Panel (b) in Figure \ref{fig:nonthermal} includes the internal energy of the preheated plasma with temperature $T_0$, we study the nonthermal particle portion against the total heated plasmas in the right-hand Panel (c). In other words, the denominator was the sum of the increment of the internal thermal plasma energy and the nonthermal particle energy, and the numerator was the same as Equation (\ref{eq:efficiency}). The increment in the internal energy of the thermal plasma was estimated as follows:
\begin{eqnarray}
  \Delta \varepsilon_{int} &=& \sum_{j=M,\kappa} n_j \left(\frac{3(T_j-T_0)}{mc^2}+\frac{K_1(mc^2/T_j)}{K_2(mc^2/T_j)}
           -\frac{K_1(mc^2/T_0)}{K_2(mc^2/T_0)} \right)  \nonumber \\
           &\simeq& \sum_{j=M,\kappa} n_j \left(\frac{T_j}{\Gamma_j - 1}-\frac{T_0}{\Gamma_0 - 1} \right),
\end{eqnarray}
where $K_{1,2}$ is the modified Bessel function, and $\Gamma_j$ and $\Gamma_0$ are the adiabatic indices with $\Gamma_{j,0}=5/3$ for non-relativistic temperature and $\Gamma_{j,0}=4/3$ for relativistic temperature, respectively. $\varepsilon'_{\rm ene}$ is expressed as
\begin{equation}
  \varepsilon'_{\rm ene}=
  \int_1^{\infty} (\gamma -1) (N_{\kappa}(\gamma)-N_{\kappa}^{\rm M}(\gamma)) d \gamma /
  (\Delta \varepsilon_{int}+\int_1^{\infty} (\gamma -1) (N_{\kappa}(\gamma)-N_{\kappa}^{\rm M}(\gamma)) d \gamma )
\end{equation}

Panels (b) and (c) show that the difference between $\varepsilon_{\rm ene}$ and $\varepsilon'_{\rm ene}$ was small and that almost all magnetic energy dissipated during magnetic reconnection was converted into nonthermal particle production for a relativistic plasma $T_0/mc^2 > 1$ and a weak guide magnetic field $B_G/B_0 \sim 0$.  

\section{Discussions and Conclusion}
In this paper, we investigated the energy partition between the thermal and nonthermal populations in magnetic reconnection for a pair plasma through PIC simulations, and found that most of the dissipated magnetic field energy was converted into thermal plasma heating rather than nonthermal particle acceleration for non-relativistic plasma; however, nonthermal particle production was efficient for relativistic plasma. In addition, the guide field effect, which is another important factor for controlling particle acceleration, was studied. We found that the efficiencies of both thermal heating and nonthermal accelerations were reduced by increasing the guide magnetic field. However, the efficiency was slightly enhanced in the case of a weak guide magnetic field.  It is useful to mention that the similar result has been recently obtained by \cite{French22}, in which the stronger guide fields suppress acceleration efficiency, and increase the power-law index and the injection energy defined by the transition energy from the thermal distribution to the power-law distribution. 

In the paper by \cite{Zenitani01}, the production of nonthermal particles without a guide magnetic field has already been discussed by focusing on particle acceleration around the magnetic diffusion region, where the magnetic field was weak and the particle was quickly accelerated along the inductive electric field perpendicular to the reconnection plane. The acceleration rate in the diffusion region was estimated as $d\epsilon/dt = e E v$, 
where $\epsilon = mc^2/\sqrt{1-v^2/c^2}$, $E$, and $v$ are the particle energy, inductive electric field, and particle velocity, respectively. These particles were first accelerated in the $z$ direction and then ejected by the Lorentz force of the reconnection magnetic field $B_n$. Therefore, the particles remained in the diffusion region during the gyro-period of the reconnecting magnetic field. The loss rate of the accelerated particles is described as 
$d {\rm ln} N(\epsilon)/dt = - (mc^2/\epsilon) (e B_n/mc), $ 
where $N(\epsilon)$ denotes the number of particles with energy $\epsilon$.  The right-hand side represents the gyro-frequency with relativistic energy $\epsilon$ because the particle gained energy from the electric field $E$ during the Speiser motion \citep{Speiser65,Hoshino01}. The balance between the abovementioned acceleration rate and loss rate yields
\begin{eqnarray}
  N(\epsilon) & \propto & \exp \left(-2 \frac{B_n}{E}  {\rm sinh}^{-1} \left( \sqrt{\frac{\epsilon/mc^2-1}{2}} \right) \right), \\
  \nonumber \\
  & \propto & \left \{
  \begin{array}{@{\,}lll}
    \exp \left( - \frac{B_n}{E} \frac{v}{c} \right) & \mbox{ for $\epsilon/mc^2 \sim 1$} , \\
    \\
    (\epsilon/mc^2)^{-B_n/E} & \mbox{ for $\epsilon /mc^2 \gg 1$} .
  \end{array}
  \right .
\end{eqnarray}
From the above simple analysis of the anti-parallel magnetic field topology, we found that the nonthermal power-law spectrum was formed by the relativistic inertia effect. In other words, the higher-energetic particles were effectively trapped in the diffusion region owing to the larger gyro-radius, and the acceleration lasted longer in the diffusion region. The power-law index was estimated using the ratio between the reconnecting magnetic field $B_n$ and the inductive electric field $E$. Because the relativistic reconnection satisfied the condition $E + (V_A/c) B_n \simeq 0$, the ratio of $B_n/E$ was approximately $c/V_A \sim 1$, suggesting the formation of a hard-energy spectrum.  For nonrelativistic reconnection, the spectrum becomes soft due to a large value of $B_n/E$, and the supra-thermal energy spectrum may be modified from a power-law function to an $\exp$-type. 

For a finite guide field, the accelerating particles may have a tendency to be trapped in the magnetic diffusion region by the partial magnetization effect of the guide magnetic field; thus, we can expect improvement in the acceleration efficiency. This mechanism may explain the improvement in the efficiency of nonthermal particle acceleration for the weak guide field cases shown in Figure \ref{fig:nonthermal}. However, for a strong guide field, we need to consider that the inductive electric field becomes weaker.
To understand the effect of the electric field on particle acceleration, we investigate the magnitude of the electric field during the active time stage of reconnection averaged downstream of the magnetic separatrix using our simulation results.

\begin{figure*}
\includegraphics[width=18cm]{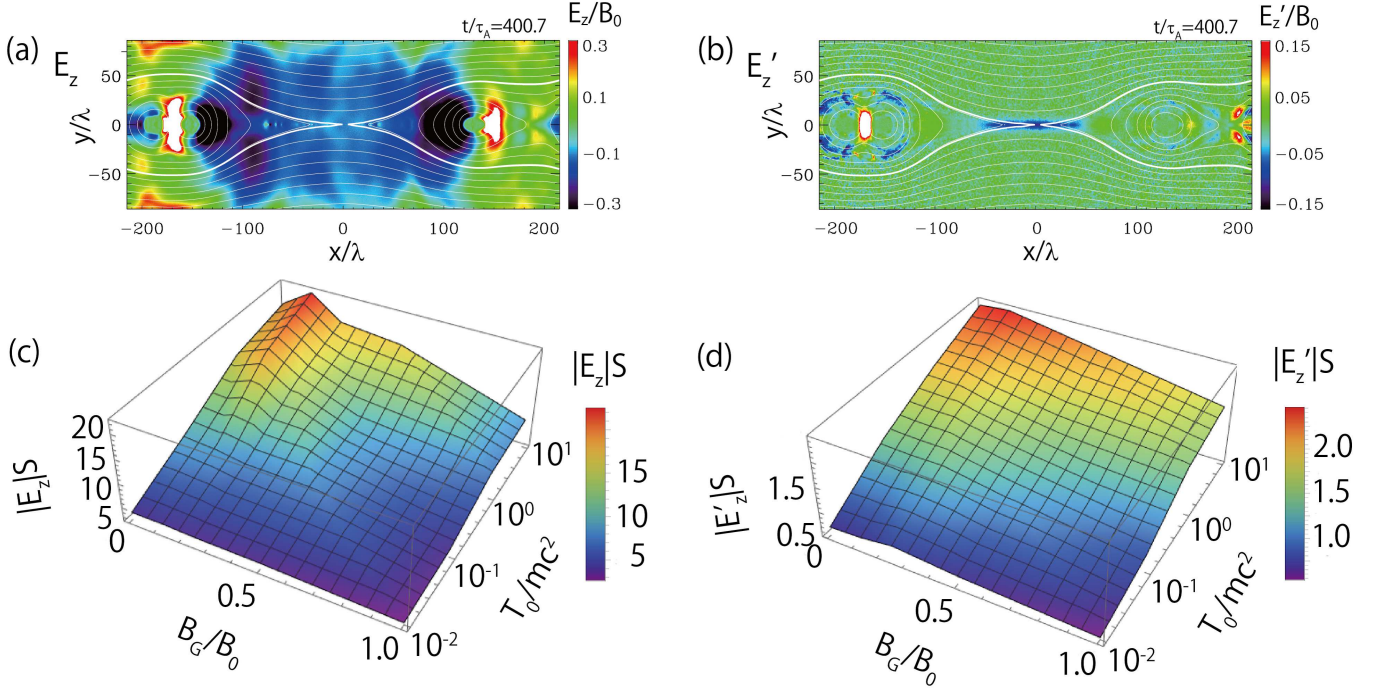}
\caption{z-component of electric fields for $E$ and $E'=E + (v/c) \times B$. The top Panels (a) and (b) show the colored contour for $T_0/mc^2=1$ and $B_G/B_0=1/2$. The bottom Panels (c) and (d) show the maximum electric fields integrated over the downstream of the separatrix as a function of $T_0/mc^2$ and $B_G/B_0$. }
\label{fig:Efield}
\end{figure*}

Figure \ref{fig:Efield} shows the electric fields $E_z$ (left) and $E'_z = E_z + ((\vec{v}/c) \times \vec{B})_z$ (right) perpendicular to the reconnection plane. $E'_z$ represents a non-MHD (Magnetohydrodynamics) effect, and $E'_z=0$ represents an ideal MHD process.
The top two panels (a) and (b) show the colored contour plots in the $x-y$ plane for $T_0/mc^2=1$ and $B_G/B_0=1/2$, and the bottom panels (c) and (d) show the maximum magnitudes of the electric field integrated over the downstream region as a function of plasma sheet temperature $T_0/mc^2$ and guide magnetic field $B_G/B_0$, as described below.
\begin{displaymath}
  |E_z|S = {\rm max}\left( \int_{\rm dw(t)} \frac{|E_z(t,x,y)| dx dy}{B_0 L_x \lambda} \right),
\end{displaymath}
where ${\rm dw(t)}$ indicates the downstream region as the function of time. 
$|E'_z|S$ has a similar definition. We normalize the electric field $|E_z|$ using the Harris magnetic field $B_0$, which is regarded as a free energy source, and the area of the downstream region is normalized by $L_x \lambda$. We adopted a maximum value of $\int |E_z(x,y,t)|dxdy$ for the time evolution of magnetic reconnection.  The top panels are not necessarily the time stage with the maximum electric field; however, these plots show a profile similar to that of the stage with the maximum value.  

Although the electric field $E_z$ in Panel (a) indicates a widely spread negative (bluish) region around the X-type point in association with the convection motion of $v \times B$, the non-MHD electric field $E'_z$ in Panel (b) is localized around the X-type point at $x/\lambda=0$, suggesting a magnetic diffusion region. The negative (reddish and white) region around $x/\lambda \sim -180$ is another magnetic diffusion region where the two plasmoids coalesce. Because the time scale of the coalescence was fast \citep{Pritchett79}, the magnitude of $E'_z$ around $x/\lambda \sim -180$ was larger than that around $x/\lambda =0$ \citep[for example, ][]{Oka10}.

Considering the magnitudes of the electric field of $|E_z|S$ and $|E'_z|S$ in Panels (c) and (d),
as expected, we find that the electric field increases with increasing plasma sheet temperature $T_0/mc^2$. 
The flow speed of the reconnection jet in the downstream is in the order of Alfv\'{e}n speed, regardless of the background plasma temperature, and the reconnection electric field $v \times B/c$ increases with increasing background plasma temperature, that is, the electric field in the reconnection region increases for a relativistic reconnection.  

When the guide field was strong and the particles were magnetized, the growth rate of reconnection was suppressed because the effective conductivity around the X-type region was large \citep{Drake77,Galeev77,Quest81,Hoshino87}. The magnetized particles remained in the diffusion region can be freely accelerated along the guide magnetic field by the action of the electric field. Subsequently, the magnetic reconnection tended to decrease the growth rate with an increase in the guide magnetic field $B_G/B_0$.  The linear growth rate was estimated as follows:
\begin{equation}
  \gamma \tau_A \sim  (k \lambda) \left( \frac{r_g}{\lambda}\right)^{3/2} \frac{\sqrt{r_g \lambda}}{l_s}
   \sim  (k \lambda) \left( \frac{r_g}{\lambda}\right)^2 \frac{B_0}{B_G},
\label{eq:growth2}
\end{equation}
where $l_s$ is the shear length of the magnetic field, defined by $l_s = ((dB_x(y)/dy)/B_G)^{-1}$.
Therefore, our result, in which the amplitude of the electric field decreases with increasing guide magnetic field, is consistent with collisionless reconnection under the guide field effect. Although the abovementioned linear growth rate only represented the early evolution of reconnection, it could help infer the magnitude of the inductive electric field during the nonlinear stage.

Comparing the profiles of $|E_z|S$ and $|E'_z|S$ in Panels (c) and (d) in Figure \ref{fig:Efield} with the efficiency of nonthermal particles $\varepsilon_{\rm ene}$ and $\varepsilon'_{\rm ene}$ shown in Panels (b) and (c) in Figure \ref{fig:nonthermal}, we find that these profiles show similar behavior, in which the larger electric field produced more energetic particles. It is difficult to determine which electric fields ($E_z$ or $E'_z$) contributed to the generation of nonthermal particles by comparing the profiles; however, the profile of $|E'_z|S$ exhibits a few similarities: a slight enhancement around $T_0/mc^2 \sim 10^{-2}$ and $B_G/B_0 \sim 0.25$ and a gradual slope against $B_G$ around $T_0/mc^2 \sim 10$. Our simple modeling of the abovementioned formation of the nonthermal particle assumed that $E'_z$ played an important role \citep{Zenitani01}.

Thus far, we studied the efficiency of nonthermal particle acceleration as a function of both the initial plasma temperature and the guide magnetic field by assuming the initial background density $N_b/N_0=5 \%$. In addition, we assumed that the background plasma temperature was the same as the initial plasma-sheet temperature $T_0$. However, the reconnection rate was controlled by the velocity of Alfv\'{e}n in the inflow region, which is described as $v_A=c/\sqrt{1+4 \pi(e+p)/B_0^2}$, where $e$ and $p$ are the total energy density and gas pressure in the inflow region, respectively.  In our previous paper \cite{Hoshino22} on anti-parallel magnetic field reconnection, we briefly discussed that energy spectrum becomes harder as the background density decreases; however, the efficiency of the nonthermal energy density did not change significantly. We expected that the basic behavior of the energy partition does not change qualitatively; however, it is important to understand these differences quantitatively in future work.

In this study, we assumed electron and positron plasmas for simplicity.  The nonthermal production should be modified for hadronic ion-electron plasmas. The power-law indices in the energy spectra observed in the solar atmosphere and in the Earth's magnetosphere are approximately $4 \sim 6$ \citep[for example][]{Oieroset2002,Lin2003}, and it appears that the efficiency of nonthermal particles is not necessarily high for non-relativistic plasmas. PIC simulation studies for non-relativistic ion-electron plasmas indicate that whereas non-thermal electrons can be quickly accelerated during reconnection \citep[for example][]{Hoshino01,Pritchett01,Drake06,Oka10}, preferential ion heating occurs for thermal population \citep{Haggerty15,Hoshino18}. The ion acceleration in nonrelativistic reconnection remains to be solved, and a large-scale simulation study is probably necessary to capture both the phenomena of electron and ion acceleration.
In astrophysical phenomena such as those observed in accretion disc coronae in black holes \citep{Remillard06,Done07} and in blazar jets \citep{Madejski16}, the hadronic ion-electron relativistic plasmas are believed to dominate instead of positron-electron pair plasmas, and the energy partition between ions and electrons is an important issue for understanding emission models that can be compared with observations.  \cite{Rowan17} and \cite{Werner18} studied plasma heating and particle acceleration on trans-relativistic reconnection for anti-parallel magnetic field, and demonstrated that energy deposition into ions is more efficient than electrons. Notably, a high ion-to-electron temperature was obtained for both nonrelativistic and trans-relativistic reconnections. However, a consistent study of the energy partition between thermal and nonthermal plasmas through nonrelativistic to relativistic plasmas remains unresolved. The energy partitioning of the ion and electron plasmas will be investigated in a separate study.

This work was supported by JSPS Grant-in-Aid for Scientific Research (KAKENHI) (Grant no. 20K20908). The author would like to thank
S. Zenitani,  S. Totorica, S. Imada, T. Amano, Y. Ohira and K. Keika for their valuable discussion.



\begin{thebibliography}{}
\bibitem[Ball et al.(2018)]{Ball18} Ball D., Sironi L., {\"O}zel F., 2018, ApJ, 862, 80. doi:10.3847/1538-4357/aac820
\bibitem[Bell (1978)]{Bell78} Bell, A.~R.\ 1978, Monthly Notices of the Royal Astronomical Society, 182, 147
\bibitem[Bessho \& Bhattacharjee(2012)]{Bessho12} Bessho, N. \& Bhattacharjee, A.\ 2012, \apj, 750, 129. doi:10.1088/0004-637X/750/2/129
\bibitem[Birn \& Priest(2007)]{Birn07} Birn, J. \& Priest, E.~R.\ 2007, Reconnection of magnetic fields : magnetohydrodynamics and collisionless theory and observations / edited by J. Birn and E. R. Priest. Cambridge : Cambridge University Press, 2007. ISBN: 9780521854207 (hbk.)
\bibitem[Blandford and Ostriker (1978)]{Blandford78} Blandford, R.~D., \& Ostriker, J.~P.\ 1978, The Astrophysical Journal, 221, L29
\bibitem[Blandford et al.(2017)]{Blandford17} Blandford, R., Yuan, Y., Hoshino, M., \& Sironi, L.,\ 2017, Space Science Reviews, 207, 291
\bibitem[Cerutti et al.(2012a)]{Cerutti12a} Cerutti, B., Uzdensky, D.~A., \& Begelman, M.~C.\ 2012, \apj, 746, 148. doi:10.1088/0004-637X/746/2/148
\bibitem[Cerutti et al.(2012b)]{Cerutti12b} Cerutti, B., Werner, G.~R., Uzdensky, D.~A., et al.\ 2012, \apjl, 754, L33. doi:10.1088/2041-8205/754/2/L33
\bibitem[Cerutti et al.(2013)]{Cerutti13} Cerutti, B., Werner, G.~R., Uzdensky, D.~A., et al.\ 2013, \apj, 770, 147. doi:10.1088/0004-637X/770/2/147
\bibitem[Cerutti et al.(2014)]{Cerutti14} Cerutti, B., Werner, G.~R., Uzdensky, D.~A., et al.\ 2014, \apj, 782, 104. doi:10.1088/0004-637X/782/2/104
\bibitem[Done et al.(2007)]{Done07} Done, C., Gierli{\'n}ski, M., \& Kubota, A.\ 2007, \aapr, 15, 1. doi:10.1007/s00159-007-0006-1
\bibitem[Drake \& Lee(1977)]{Drake77} Drake, J.~F. \& Lee, Y.~C.\ 1977, Physics of Fluids, 20, 1341. doi:10.1063/1.862017
\bibitem[Drake et al.(2006)]{Drake06} Drake, J.~F., Swisdak, M., Che, H., et al.\ 2006, \nat, 443, 553. doi:10.1038/nature05116
\bibitem[French et al.(2022)]{French22} French, O., Guo, F., Zhang, Q., et al.\ 2022, arXiv:2210.08358. doi:10.48550/arXiv.2210.08358
\bibitem[Galeev \& Zeleny{\v{i}}(1977)]{Galeev77} Galeev, A.~A. \& Zeleny{\v{i}}, L.~M.\ 1977, Soviet Journal of Experimental and Theoretical Physics Letters, 25, 380
\bibitem[Guo et al.(2014)]{Guo14} Guo, F., Li, H., Daughton, W., \& Liu, Y.-H.\ 2014, Physical Review Letters, 113, 155005
\bibitem[Guo et al.(2020)]{Guo20} Guo, F., Liu, Y.-H., Li, X., Li, H., Daughton, W., and Kilian, P.\ 2020, Physics of Plasmas, 27, 080501. doi:10.1063/5.0012094
\bibitem[Haggerty et al.(2015)]{Haggerty15} Haggerty, C.~C., Shay, M.~A., Drake, J.~F., et al.\ 2015, \grl, 42, 9657. doi:10.1002/2015GL065961
\bibitem[Harris(1962)]{Harris62} Harris, E.~G.\ 1962, Il Nuovo Cimento, 23, 115
\bibitem[Hoshino(1987)]{Hoshino87} Hoshino, M.\ 1987, \jgr, 92, 7368. doi:10.1029/JA092iA07p07368
\bibitem[Hoshino et al.(2001)]{Hoshino01} Hoshino, M., Mukai, T., Terasawa, T., \& Shinohara, I.\ 2001, \jgr, 106, 25979. doi:10.1029/2001JA900052
\bibitem[Hoshino \& Lyubarsky(2012)]{Hoshino12} Hoshino, M., \& Lyubarsky, Y.\ 2012, Space Science Reviews, 173, 521
\bibitem[Hoshino(2015)]{Hoshino15} Hoshino, M.\ 2015, \prl, 114, 061101. doi:10.1103/PhysRevLett.114.061101
\bibitem[Hoshino(2018)]{Hoshino18} Hoshino, M.\ 2018, \apjl, 868, L18. doi:10.3847/2041-8213/aaef3a
\bibitem[Hoshino(2022)]{Hoshino22} Hoshino, M.\ 2022, Physics of Plasmas, 29, 042902. doi:10.1063/5.0086316
\bibitem[Jaroschek et al.(2004)]{Jaroschek04} Jaroschek, C.~H., Lesch, H., \& Treumann, R.~A.\ 2004, The Astrophysical Journal, 605, L9
\bibitem[Jaroschek \& Hoshino(2009)]{Jaroschek09} Jaroschek, C.~H. \& Hoshino, M.\ 2009, \prl, 103, 075002. doi:10.1103/PhysRevLett.103.075002
\bibitem[Kirk(2004)]{Kirk04} Kirk, J.~G.\ 2004, \prl, 92, 181101. doi:10.1103/PhysRevLett.92.181101
\bibitem[Lin et al.(2003)]{Lin2003} Lin, R.~P., Krucker, S., Hurford, G.~J., et al.\ 2003, \apjl, 595, L69. doi:10.1086/378932
\bibitem[Liu et al.(2011)]{Liu11} Liu, W., Li, H., Yin, L., et al.\ 2011, Physics of Plasmas, 18, 052105. doi:10.1063/1.3589304
\bibitem[Lyutikov(2003)]{Lyutikov03} Lyutikov, M.\ 2003, \mnras, 346, 540. doi:10.1046/j.1365-2966.2003.07110.x
\bibitem[Madejski \& Sikora(2016)]{Madejski16} Madejski, G. (Greg) . \& Sikora, M.\ 2016, \araa, 54, 725. doi:10.1146/annurev-astro-081913-040044
\bibitem[{\O}ieroset et al.(2002)]{Oieroset2002} {\O}ieroset, M., Lin, R.~P., Phan, T.~D., et al.\ 2002, \prl, 89, 195001. doi:10.1103/PhysRevLett.89.195001
\bibitem[Oka et al.(2010)]{Oka10} Oka, M., Phan, T.-D., Krucker, S., et al.\ 2010, \apj, 714, 915. doi:10.1088/0004-637X/714/1/915
\bibitem[Oka et al.(2015)]{Oka15} Oka, M., Krucker, S., Hudson, H.~S., et al.\ 2015, \apj, 799, 129. doi:10.1088/0004-637X/799/2/129
\bibitem[Petropoulou \& Sironi(2018)]{Petropoulou18} Petropoulou, M. \& Sironi, L.\ 2018, \mnras, 481, 5687. doi:10.1093/mnras/sty2702
\bibitem[Pritchett \& Wu(1979)]{Pritchett79} Pritchett, P.~L. \& Wu, C.~C.\ 1979, Physics of Fluids, 22, 2140. doi:10.1063/1.862507
\bibitem[Pritchett(2001)]{Pritchett01} Pritchett, P.~L.\ 2001, \jgr, 106, 3783. doi:10.1029/1999JA001006
\bibitem[Quest \& Coroniti(1981)]{Quest81} Quest, K.~B. \& Coroniti, F.~V.\ 1981, \jgr, 86, 3289. doi:10.1029/JA086iA05p03289
\bibitem[Remillard \& McClintock(2006)]{Remillard06} Remillard, R.~A. \& McClintock, J.~E.\ 2006, \araa, 44, 49. doi:10.1146/annurev.astro.44.051905.092532
\bibitem[Rowan et al.(2017)]{Rowan17} Rowan, M.~E., Sironi, L., \& Narayan, R.\ 2017, \apj, 850, 29. doi:10.3847/1538-4357/aa9380
\bibitem[Sironi \& Spitkovsky(2011)]{Sironi11} Sironi, L. \& Spitkovsky, A.\ 2011, \apj, 726, 75. doi:10.1088/0004-637X/726/2/75
\bibitem[Sironi et al.(2014)]{Sironi14} Sironi, L., \& Spitkovsky, A.\ 2014, The Astrophysical Journal, 783, L21
\bibitem[Speiser(1965)]{Speiser65} Speiser, T.~W.\ 1965, \jgr, 70, 4219. doi:10.1029/JZ070i017p04219
\bibitem[Uzdensky(2016)]{Uzdensky16} Uzdensky, D.~A.\ 2016, Magnetic Reconnection: Concepts and Applications, 473
\bibitem[Vasyliunas(1968)]{Vasyliunas68} Vasyliunas, V.~M.\ 1968, \jgr, 73, 2839. doi:10.1029/JA073i009p02839
\bibitem[Werner et al.(2016)]{Werner16} Werner, G.R., Uzdensky, D.~A., Cerutti, B., Nalewajko, K., \& Begelman, M.~C.\ 2016, \apjl, 816, L8. doi:10.3847/2041-8205/816/1/L8
\bibitem[Werner et al.(2018)]{Werner18} Werner, G.~R., Uzdensky, D.~A., Begelman, M.~C., et al.\ 2018, \mnras, 473, 4840. doi:10.1093/mnras/stx2530
\bibitem[Zenitani \& Hoshino(2001)]{Zenitani01} Zenitani, S., \& Hoshino, M.\ 2001, The Astrophysical Journal, 562, L63
\bibitem[Zenitani \& Hoshino(2005a)]{Zenitani05a} Zenitani, S. \& Hoshino, M.\ 2005, \apjl, 618, L111. doi:10.1086/427873  
\bibitem[Zenitani \& Hoshino(2005b)]{Zenitani05b} Zenitani, S. \& Hoshino, M.\ 2005, \prl, 95, 095001. doi:10.1103/PhysRevLett.95.095001
\bibitem[Zenitani \& Hoshino(2007)]{Zenitani07} Zenitani, S. \& Hoshino, M.\ 2007, \apj, 670, 702. doi:10.1086/522226
\bibitem[Zenitani \& Hoshino(2008)]{Zenitani08} Zenitani, S. \& Hoshino, M.\ 2008, \apj, 677, 530. doi:10.1086/528708
\bibitem[Zweibel \& Yamada(2009)]{Zweibel09} Zweibel, E.~G. \& Yamada, M.\ 2009, \araa, 47, 291. doi:10.1146/annurev-astro-082708-101726
\end{thebibliography}
\end{document}